# Thermoplasmonic Effect of Surface Enhanced Infrared Absorption in Vertical Nanoantenna Arrays

Andrea Mancini†, Valeria Giliberti‡, Alessandro Alabastri§, Eugenio Calandrini∥, Francesco De Angelis∥, Denis Garoli∥* and Michele Ortolani†*

† Dipartimento di Fisica, Sapienza Università di Roma, Piazzale Aldo Moro 5, 00185 Rome, Italy

‡ Center for Life NanoSciences, Istituto Italiano di Tecnologia (IIT), Viale Regine Elena 291, 00185 Rome, Italy

§ Department of Physics and Astronomy and Department of Electrical and Computer Engineering, Rice University, 6100 Main Street, Houston, Texas 77005, United States

∥ Nanostructures Department, Istituto Italiano di Tecnologia (IIT), 16163 Genova, Italy

*corresponding author e-mails: denis.garoli@iit.it , michele.ortolani@roma1.infn.it

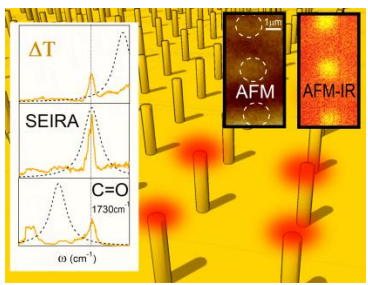

**Abstract.** The temperature increase and temperature gradients induced by mid-infrared laser illumination of vertical gold nanoantenna arrays embedded into polymer layers was measured directly with a photothermal expansion nanoscope. Nanoscale thermal hotspot images and local temperature increase spectra were both obtained, the latter by broadly tuning the emission wavelength of a quantum cascade laser. The spectral analysis indicates that plasmon-enhanced mid-infrared vibrations of molecules located in the antenna hotspots are responsible for some of the thermoplasmonic resonances, while Joule heating in gold is responsible for the remaining resonances. In particular, plasmonic dark modes with low scattering cross-section mostly produce surface-enhanced infrared absorption (SEIRA), while bright modes with strong radiation coupling produce Joule heating. The dark modes do not modify the molecular absorption lineshape and the related temperature increase is chemically triggered by the presence of molecules with vibrational fingerprints resonant with the plasmonic dark modes. The bright modes, instead, are prone to Fano interference, display an asymmetric molecular absorption lineshape and generate heat also at frequencies far from molecular vibrations, insofar lacking chemical specificity. For focused mid-infrared laser power of 50 mW, the measured nanoscale temperature increases are in the range of 10 K and temperature gradients reach 5 K/µm in the case of dark modes resonating with strong infrared vibrations such as the C=O bond of poly-methylmethacrylate at 1730 cm$^{-1}$.

Thermoplasmonics has emerged as a new way of increasing temperature remotely by light beams [1, 2], exploiting the Joule heating effect at optical frequencies in metal nanoparticles [3-5] and nanoantennas [6-8]. Thermoplasmonics has been explored up to now with two different aims: (1) to produce a strong, local temperature increase at the nanoscale, in close proximity of single plasmonic nanoparticles in which light excitation results in high local currents hence strong power dissipation in the metal [4-7, 9-12]; or (2) to increase the efficiency of radiation heating processes in large volumes or surfaces filled with both target molecules to be heated and plasmonic nanostructures acting as mediators of electromagnetic energy absorption for the entire system (global temperature increase) [3, 8, 12]. The application of thermoplasmonic concepts has been carried out mostly with plasmonic nanoparticles randomly dispersed in solutions [3,4] or with single plasmonic antennas [6, 7], however antenna arrays have also been considered [14], because spatially coherent surface plasmon effects in periodic arrays may further enhance the absorption efficiency [15]. Remotely light-activated local or global temperature increases can be of extreme importance in e.g. catalysis [16, 17], medical therapy [12], material synthesis [10, 11], magnetic assisted recording [6], triggering of phase transitions [4, 9], and thermophoresis [18].

The transient electron temperature at the metal surface of antennas in the field-enhancement regions can be far higher than the steady-state metal lattice temperature [1], leading to so-called hot electron effects such as electron tunneling emission [16, 17, 19]. Hot electron effects are distinct from Joule heating effect, although they may ultimately contribute to local and global heating. In the present work, we introduce a novel path to thermoplasmonics based on electromagnetic (e.m.) energy dissipation by non-radiative decay of enhanced molecular dipole vibrations in the field-enhancement hotspots adjacent to the antennas, which one could broadly consider as the counterparts in the dielectric half-space of the hot electron effects in the metal half-space. This mechanism of temperature increase is activated by laser illumination at substance-specific vibrational fingerprints in the mid-infrared (IR) range and belongs to the class of phenomena called surface-enhanced infrared absorption (SEIRA) [20, 21]. SEIRA has seldom been exploited up to now for thermoplasmonic applications due to the lack of practical tunable high-power mid-IR semiconductor light sources, but these have now become commercially available after the advent of external-cavity tunable quantum cascade lasers (EC-QCLs) [22].

The strength of IR vibrational dipoles of virtually all organic molecules is huge if compared to any other absorption of e.m. radiation in the infrared and visible ranges, except maybe some dyes. Also, vibration typically occurs at molecule-specific frequencies in the "IR fingerprint" range 1000-2000 $cm^{-1}$ (wavelengths between 5 and 10 $\mu$m). Nanoscale mid-IR thermoplasmonics with tunable EC-QCLs has been recently exploited for mid-IR absorption nanospectroscopy of few molecules [23-25], nonlinear optics in the mid-IR based on phase transitions of liquid crystals [26], and mapping of field-enhancement hotspots in IR metamaterials [27-29]. SEIRA has been long sought for in mid-IR plasmonic nanoantenna structures, however it has been elusive up to now due to Fano interference phenomena [30-36] that can prevent IR absorption enhancement while providing scattering enhancement [37, 38]. In this work, we have observed true SEIRA lineshapes (i.e. Lorentzian lineshapes of enhanced intensity centered at the molecule vibration frequency) in plasmonic nanoantenna arrays by exploiting a scanning probe tip-enhanced mid-IR spectroscopy technique that measures directly the local temperature increase $\Delta T$ with nanoscale resolution. A key factor to reduce interference effects preventing the experimental observation of SEIRA in nanoantennas was the exploitation of plasmonic "dark" modes of the array, featuring small scattering cross sections. The dark modes display a moderate field enhancement that is sufficient to enhance the vibrations of molecules in the hotspots, but they do not strongly interact with free-space radiation, leading to reduced Fano interference and smaller Joule heating effects if compared to bright modes.

**Experimental Section**

The vertical antennas consist in protrusions made of photoresist polymer spin-coated on a silicon wafer and then hardened by exposure to a focused ion beam. The non-exposed part of the film is removed in acetone. The resulting protrusions have a diameter of ~200 nm and a height $H$ equal to the initial polymer film thickness. The fabrication process is entirely described in Ref. [39]. The antenna arrays, including the flat silicon wafer surface, were coated conformally with a 80 nm thick evaporated gold film, closing the access to the hollow cavity inside the protrusion and leading to a final gold-coated vertical rod antenna structure with diameter $d_{\text{ant}} \sim 360$ nm [39, 40]. The samples used in this work are square periodic arrays of vertical nanoantennas with $H$ ranging from 2.2 to 2.7 $\mu$m and pitch $P$ equal to 3.0, 3.5 or 4.0 $\mu$m. After fabrication, the vertical antennas on each sample were embedded up to their top in a spin-coated polymer bilayer film, so as to allow their nanoscale investigation by atomic force microscopy (AFM). The entire body of the antennas was embedded in a weak IR-absorbing polymer (AZ5214, spin-coating speed 1000 to 2000 rpm, final thickness 3.1 to 2.1 $\mu$m), while a thin layer of a strong IR absorbing polymer was spin-coated on top of the vertical antennas (PMMA 950k, 2% solids in ethyl-lactate, spin-coating speed 3000 to 2000 rpm, final thickness 100 to 400 nm) so that to fill the region where the antenna hotspots are expected.

The mid-IR far-field response of the arrays is measured with a commercial Fourier-transform infrared (FTIR) spectrometer (*Bruker IFS66v/S*) coupled to a reflective microscope (*Bruker Hyperion*). Unpolarized radiation is focused with a Cassegrain objective (incidence angle range of 10° to 30° with respect to the surface normal) onto a 70×70 $\mu$m² square spot at the sample surface with an direction. A gold mirror is used as reference to compute the absolute reflectance $R(\omega)$ and the extinction is calculated as $E(\omega) = 1 - R(\omega)$ because transmission is zero. Each spectrum is obtained as the average of 1024 interferometer scans at 4 cm$^{-1}$ spectral resolution and the spectral range is 650 to 3000 cm$^{-1}$.

The nanoscale thermoplasmonic response of the system is investigated with an AFM operating in contact mode with the mechanical resonance-enhanced photothermal expansion technique (AFM-IR, *Anasys Instruments NanoIR2* with top side illumination). The mid-IR light source is a tunable EC-QCL (*Daylight Solutions MIRcat-PX-B*), with continuous wavelength (wavenumber) range of 5.5 to 9.1 μm (1900 to 1100 cm$^{-1}$) and accordable laser power range from 1 to 500 mW. To achieve full illumination of the probe tip, the laser beam impinges on the sample with a 70° angle with respect to the surface normal in *p*-polarization, leading to an elliptic focal spot of 30×10 $\mu m^2$ centered on the AFM probe tip. A laser power of 50 mW gives a laser power density in the focus of 1.6 10$^8$ W/m$^2$. The AFM-IR photothermal expansion maps are obtained by monitoring the AFM probe cantilever deflection oscillations at the repetition rate of the EC-QCL, which is set to match the mechanical resonance of the cantilever (here, 220 kHz). The voltage scale of the AFM-IR maps is the AFM position-sensitive photodetector signal $V_{PSPD}$ at the EC-QCL repetition rate of 220 kHz, measuring the laser-induced AFM cantilever deflection oscillations. The AFM topography maps are simultaneously recorded from the dc component of the deflection signal. The laser pulse duration is 260 ns, the duty cycle is 4%. The spectral resolution is 2 cm$^{-1}$. Silicon probe tips with slanted tip shaft (*Nanosensors AdvancedTEC*) were employed.

The e.m. calculations were performed using the commercial software COMSOL Multiphysics. Calculations were carried out by placing the antennas (described by a Drude-like electric permittivity) on a metal half-space and filling the other half-space with a background medium (refractive index 1.6) up to the antenna height and placing air (refractive index 1) elsewhere. The fabricated and experimentally characterized samples were modeled as infinite three-dimensional arrays of antennas where Floquet periodic conditions were set in the planar directions to define the rectangular unit cells. Input radiation was accounted for by setting e.m. ports at the top surface of the modeled domain, on the interior side of the perfect matching layer, setting an angle of incidence $\theta_{inc}$. The incident radiation was represented by a linearly polarized plane wave, the absorption and scattering spectra were obtained, and the field enhancement maps were then calculated at the two resonance frequencies derived from the spectra.

**Results**

In Figure 1a-d a typical sample structure is shown with SEM images taken at different view angles. The antennas are constituted by gold-coated vertical cylinders protruding from a flat surface, also gold-coated. The geometrical parameters (height *H* and array pitch *P*) were chosen on the basis of previous works [40] so as to overlap the plasmonic resonance frequencies in the mid-IR range with the molecular vibration frequencies of the two different polymers (AZ and PMMA) embedding the vertical antennas. As a further fine-tuning parameter for obtaining better frequency overlap, the vertical antennas were also fabricated with a tilt angle $\theta_{tilt} = 5°, 10°, 20°$ with respect to the surface-normal direction. The tilt angle acts as a small perturbation of the plasmonic modes of the array, whose resonance frequency is mainly set by *P* and *H* and by the incidence angle of the illuminating IR beam $\theta_{inc}$. After fabrication, the vertical antennas embedded in polymers were investigated by AFM-IR, a technique suitable for mapping the local temperature increase $\Delta T(\omega)$ in the antenna hotspots (see an example in inset of Figure 1a) [27-29, 41].

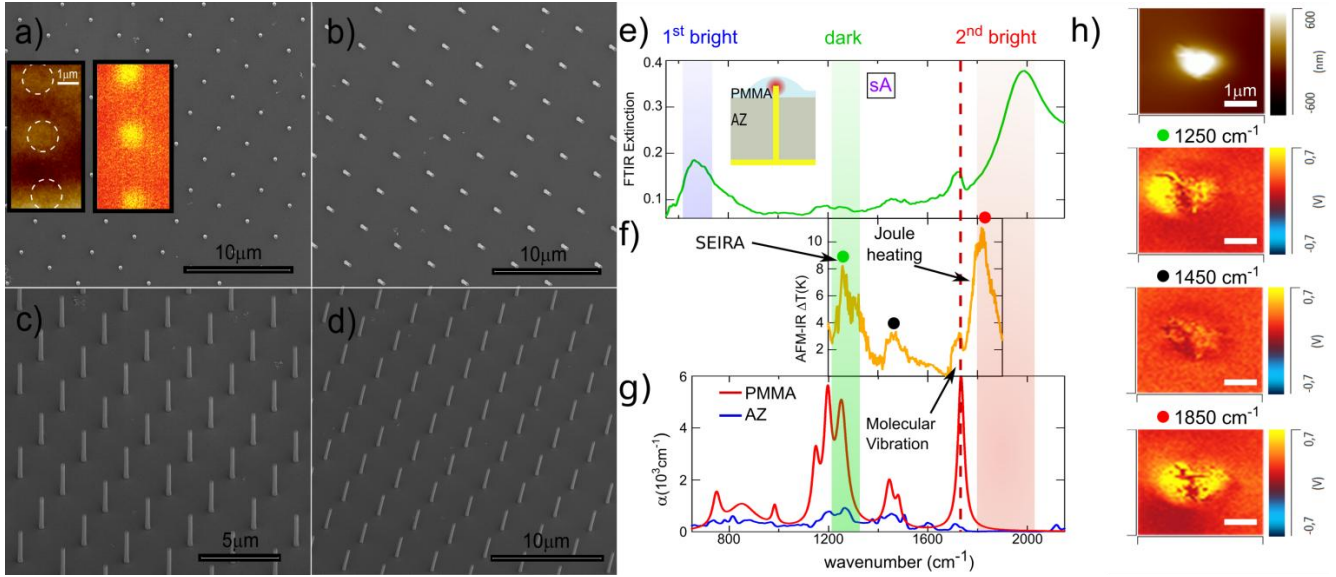

**Figure 1.** a-d: Scanning electron micrographs of an array of vertical antennas, with zero tilt angle (a,c) and with 10° tilt angle (b,d). Insets of panel a: AFM topography and the AFM-IR photothermal expansion image of a portion of the array embedded into a polymer layer. e) extinction; and f) AFM-IR spectrum of sample sA with $P = 3.5$ $\mu$m, $H = 2.7$ $\mu$m. Bright and dark array resonances are highlighted by shaded color areas. g) absorption coefficients $\alpha(\omega)$ of the embedding polymers PMMA and AZ. The 1250 cm$^{-1}$ PMMA vibration is enhanced by the dark mode (SEIRA, green dot). The red vertical line indicates the 1730 cm$^{-1}$ PMMA vibration, not enhanced by the bright mode. Joule heating vy the bright mode is observed at 1850 cm$^{-1}$ (red dot). h) Topography and AFM-IR maps of a single antenna of sample sA. The antenna position offset in the different AFM-IR maps is due to sample drift during the map acquisition time of 20 minutes.

The embedded arrays were first investigated by FTIR spectroscopy by measuring their absolute reflectance $R(\omega)$ at $\theta_{inc} = 20°$. The measured IR extinction spectra $E(\omega) = 1 - R(\omega)$ include both scattering and absorption losses, however only the absorption losses contribute to the temperature increase, while for scattering losses the e.m. energy is radiated away from the system. Plasmonic modes with high/low scattering efficiency are usually called bright/dark modes [27, 30, 31]. Indeed, it has been recognized that transmission/reflection spectroscopy is not suitable to isolate thermoplasmonic effects among other plasmonic field-enhancement effects [34]. While several techniques have been developed to directly measure the thermoplasmonic $\Delta T$ [42-45], AFM-IR features nanoscale mapping resolution for imaging the local $\Delta T$ in plasmonic resonators beyond the diffraction limit [27, 29]. AFM-IR measures the photothermal expansion of the material under a scanning probe tip, which ultimately depends on the absorption only [23-25, 27]. Furthermore, with the same AFM-IR setup a laser-induced temperature increase spectrum $\Delta T(\omega)$ can be also acquired by continuously tuning the EC-QCL wavelength with the probe tip position kept fixed. The $\Delta T(\omega)$ curves taken in different sample locations can be used to study local absorption efficiency and heat fluxes at the nanoscale. In this work, dielectric AFM scanning probes with slanted tip shaft have been employed to minimize the perturbation of the plasmonic antenna fields by the probe tip, which also does not significantly absorb mid-IR radiation.

In Figure 1e and 1f, the extinction $E(\omega)$ measured by FTIR and the local temperature increase $\Delta T(\omega)$ measured by AFM-IR are shown for a representative antenna array (sample sA) embedded in an AZ-PMMA polymer bilayer structure such that the antenna hotspot is filled with PMMA. Considering the molecular absorption coefficient spectra in Figure 1g, AZ can be considered as a weak IR absorber and PMMA as a strong IR absorber at three main molecule absorption bands around 1250, 1450 and 1730 cm$^{-1}$ corresponding to C-O-C stretching, C-H bending and C=O (carbonyl) stretching vibrations respectively. In Figure 1e, two peaks are observed in $E(\omega)$ at 650 cm$^{-1}$ and 1950 cm$^{-1}$ that can be attributed to the first and second bright plasmonic modes of the antenna array, in agreement with the e.m. simulations (see Figure 2c). In Figure 1f, a strong thermoplasmonic peak is observed in $\Delta T(\omega)$ at 1850 cm$^{-1}$, where there is no molecular vibration, and it is attributed to Joule heating of the antennas at the second bright plasmonic mode. We define $\omega_{2b}$ as the frequency at which the second bright mode displays maximum Joule heating effect in $\Delta T(\omega)$. The small shift between the peak frequency in $E(\omega)$ and the one in $\Delta T(\omega)$ is both due to the difference in resonance frequencies for far-field and near-field excitation [45, 46] and to the different illumination configurations of Figure 1e ($\theta_{inc} = 20°$) and Figure 1f ($\theta_{inc} = 70°$).

Beyond the bright mode peak at $\omega_{2b}$ = 1850 cm$^{-1}$, a second peak in $\Delta T(\omega)$ is observed in Figure 1f at 1250 cm$^{-1}$ but it has no counterpart in the extinction spectrum of Figure 1e. As already demonstrated in Ref. [36], a further plasmonic mode of vertical nanoantenna arrays emerges at intermediate wavelengths between the first and second bright modes for $\theta_{inc} \gtrsim 45°$, due to major breaking of the illumination symmetry. The purely thermoplasmonic nature of the $\Delta T(\omega)$ peaks observed at 1850 cm$^{-1}$ and at 1250 cm$^{-1}$ of Figure 1f is confirmed by the detection of antenna hotspots in the AFM-IR signal maps [27-29, 41] shown in the inset of Figure 1a and, more in details, in Figure 1h. As a control experiment, the map of Figure 1h taken at the molecular absorption frequency of 1450 cm$^{-1}$ does not show any antenna hotspot, because heat is generated everywhere in the sample. Therefore, for illumination at 1450 cm$^{-1}$ there is a global temperature increase only due to direct (Beer-Lambert) molecular absorption, but no temperature gradient due to antenna resonances. The hotspots seen in the AFM-IR maps of Figure 1h display half-width half-maximum diameters around 600 nm. In vertical nanoantennas, however, the field-enhancement region (i.e. the true e.m. hotspot), is only slightly broader than $d_{ant} \sim 360$ nm [36, 40]. The difference between the hotspot size measured with AFM-IR imaging and the true e.m. hotspot size is to be attributed to the effect of thermal diffusion in polymers [23, 24].

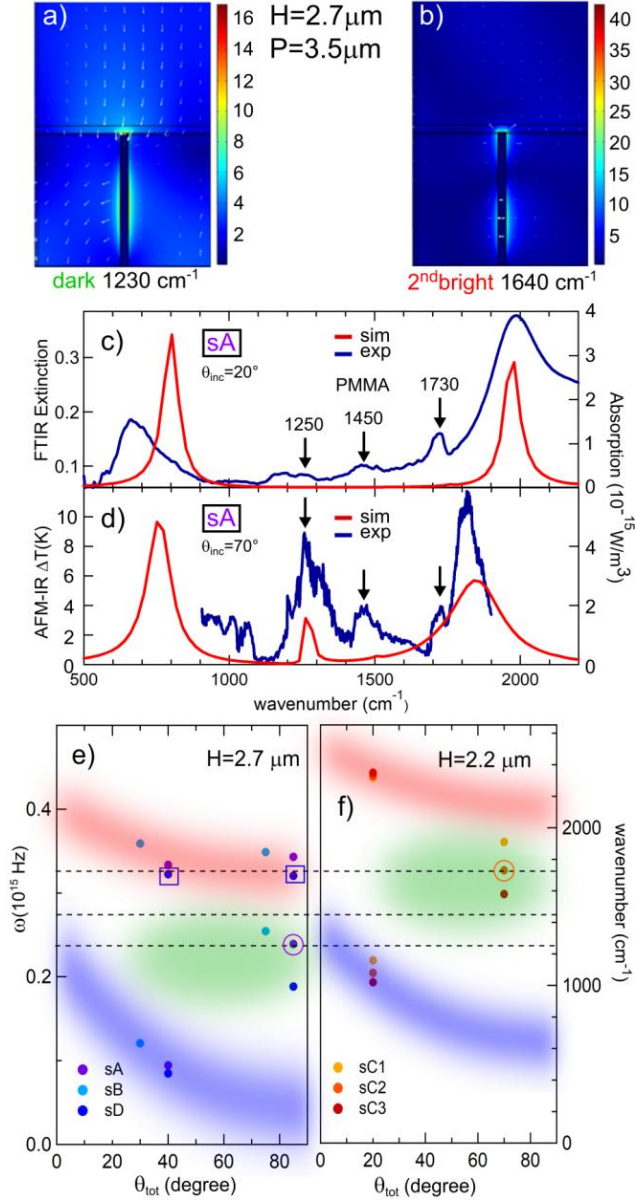

**Figure 2.** Simulated field enhancement maps (a,b) and absorption spectra (red curves in c,d, calculated for $H = 2.0$ μm, $P = 3.0$ μm). The experimental extinction and AFM-IR spectra (blue curves in c,d) show both the plasmonic modes and the imprint of molecular vibrations of PMMA (black arrows in c,d, and as dashed lines in e,f). In e,f, the measured plasmonic mode frequency as a function of $\theta_{tot}$ for all samples is shown together with shaded areas representing the expected qualitative behavior of "spoof" surface plasmon polariton dispersions. Circles/squares highlight resonance between dark/bright modes and PMMA vibrations.

In Figure 2a and 2b the simulated field-enhancement maps for the dark mode and the second bright mode of a structure with geometrical parameters identical to sample sA are reported for $\theta_{inc} = 70°$. The maps in Figure 2a and 2b confirm the existence of e.m. hotspots on top of the vertical antennas and of additional field enhancement regions along the antenna body. The molecules contained in these field enhancement regions undergo enhanced mid-IR absorption (SEIRA), which contribute to measurable $\Delta T$ at the surface locations just above the antenna. This is the case for sample sA, where the PMMA vibration at 1250 cm$^{-1}$ approximately corresponds to the dark mode frequency. In addition to SEIRA, one has the conventional thermoplasmonic Joule heating effect in the metal structure, which also heats up the polymer molecules at surface locations above the antenna. The simplified simulations shown in Figure 2c and 2d as red curves, representing the approximate mode spectra of a typical vertical antenna array for $\theta_{inc} = 20°$ and $\theta_{inc} = 70°$ respectively, compare well with $E(\omega)$ measured by FTIR and $\Delta T(\omega)$ measured by AFM-IR on sample sA (blue curves). The effect of SEIRA and the effect of Joule heating can be disentangled by studying a number of samples in which either dark or bright plasmonic modes are made to resonate with specific molecular vibrations in order to clarify the role of illumination symmetry breaking in SEIRA [27, 30, 31].

In Figure 2e and 2f the experimental plasmonic mode frequencies of all samples of Table 1 are reported vs. the angle $\theta_{tot} = \theta_{tilt} + \theta_{inc}$. The antenna height $H$ is the main characteristic length for frequency scaling, therefore samples with two different values $H$ (2.7 and 2.2 $\mu$m) are reported in two different plots. The total angle $\theta_{tot}$ approximately accounts for the quasi-dipole excitation pattern of the vertical antennas that select a specific illumination angle, relative to the vertical antenna axis, within the broad angular distribution of the optical objectives employed in the experiments. Shaded color areas in Figure 2e and 2f represent the expected qualitative behaviors of "spoof" surface plasmons of the array [40]. Samples whose plasmonic modes resonate with PMMA vibrations are highlighted by circles (resonating dark modes) or squares (resonating bright modes).

| Sample | H | $\theta_{tilt}$ | P | $\omega_{1b}$ | $\omega_{dark}$ | $\omega_{2b}$ |
|---|---|---|---|---|---|---|
| sA | 2.7 $\mu$m | 20° | 3.5 $\mu$m | 510 cm$^{-1}$ | 1270 cm$^{-1}$ | 1820 cm$^{-1}$ |
| sB (control) | 2.7 $\mu$m | 10° | 3.5 $\mu$m | 640 cm$^{-1}$ | 1350 cm$^{-1}$ | 1850 cm$^{-1}$ |
| sC1 | 2.2 $\mu$m | 0° | 3.0 $\mu$m | 1160 cm$^{-1}$ | 1910 cm$^{-1}$ | 2350 cm$^{-1}$ |
| sC2 | 2.2 $\mu$m | 0° | 3.5 $\mu$m | 1080 cm$^{-1}$ | 1730 cm$^{-1}$ | 2330 cm$^{-1}$ |
| sC3 | 2.2 $\mu$m | 0° | 4.0 $\mu$m | 1020 cm$^{-1}$ | 1580 cm$^{-1}$ | 2320 cm$^{-1}$ |
| sD | 2.7 $\mu$m | 20° | 3.0 $\mu$m | 450 cm$^{-1}$ | <1100 cm$^{-1}$ | 1700 cm$^{-1}$ |

**Table 1**. Geometrical parameters and plasmonic mode frequencies of the investigated samples.

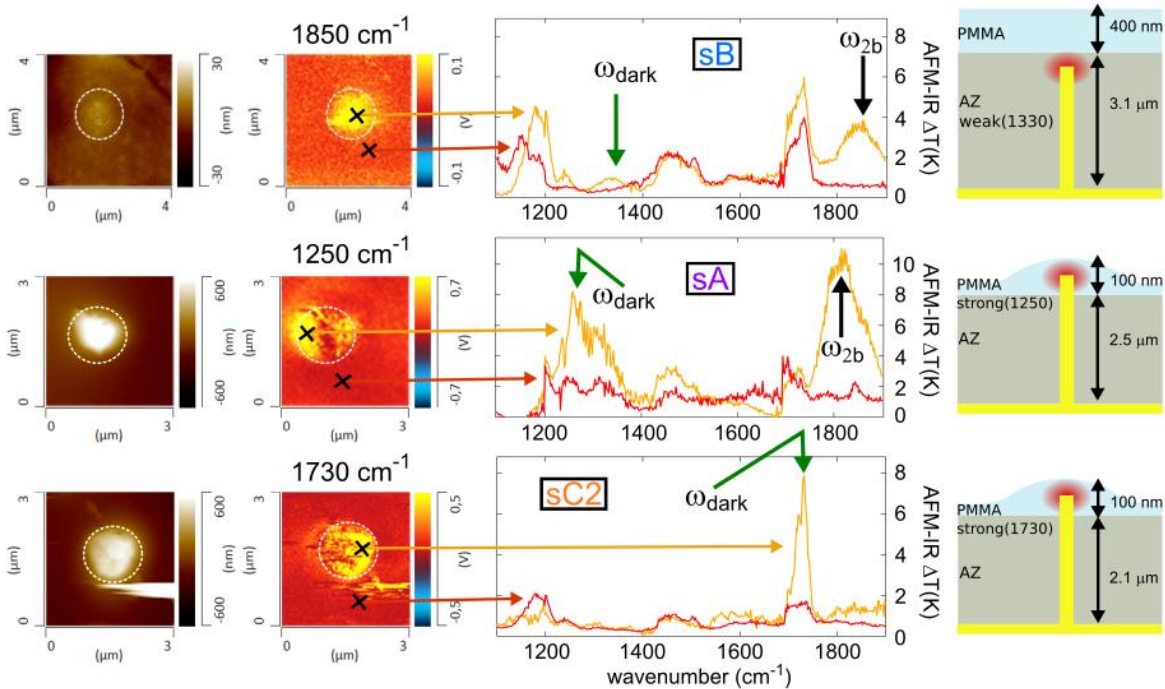

**Figure 3**. From left to right column: AFM topography map, AFM-IR photothermal expansion map, $\Delta T(\omega)$ spectra in two different locations (outside and inside the hotspots) and sketch of the embedding polymer configuration. Data for three samples are reported in the three different rows (sample parameters reported in Table 1). In the sketches on the right, the embedding polymer bilayer thicknesses are reported with the strengths and frequencies (in cm$^{-1}$) of specific molecule vibrations enhanced by the dark mode of each sample (green arrows in the spectral plots).

In Figure 3 a set of experimental $\Delta T(\omega)$ spectra and AFM-IR maps is shown for samples sB (control), sA (weak SEIRA effect) and sC2 (strong SEIRA effect). From the left to the right, one can see the AFM topography maps (sepia color scale), the corresponding AFM-IR maps (red-hot color scale), the AFM-IR spectra at two locations (one in the antenna hotspot, orange curve, and one outside the hotspot, red curve), and the sketch of the embedding polymer bilayer configuration with all thickness values. In the control sample sB, in which thicker embedding layers were used in order to displace the PMMA molecules 400 nm away from the hotspot, $\Delta T(\omega)$ in the hotspot (orange curve) displays a peak at $\omega_{2b} = 1850$ cm$^{-1}$ originating from Joule heating from the bright mode. The dark mode appears in the orange curve of sample sB only as a small peak at 1330 cm$^{-1}$, due to the absence of strong vibrational dipoles in the AZ polymer. Comparing the relative intensity of the PMMA vibrations at 1250, 1450 and 1730 cm$^{-1}$ in sample sB, one sees that they are identical for both the orange and the red curve, and also that they match their nominal ratio of 2:1:2 in the absence of plasmonic field enhancement (see Figure 1g for reference). In the $\Delta T(\omega)$ spectra of samples sA and sC2, instead, there is a resonant coupling of the plasmonic dark mode with the strong molecular vibrations of PMMA. In $\Delta T(\omega)$ of sample sA, in particular, the Joule heating peak related to the second bright mode is still present around $\omega_{2b} = 1850$ cm$^{-1}$, where no molecule vibrational frequency exists, but the dark mode clearly enhances the molecular vibrations of PMMA at 1250 cm$^{-1}$: the peak intensity ratio in the orange curve of sample sA is approximately 4:1:2, indicating a two-fold enhancement of the peak at 1250 cm$^{-1}$. In sample sC2, the bright modes lie outside the AFM-IR frequency range (see Table 1), while the dark mode is predicted by simulations around 1700 cm$^{-1}$ (see Supporting Information). In facts, from the observed peak intensity ratio of 2:1:14, the dark mode of samples sC2 produces a remarkable seven-fold enhancement of the PMMA vibration at 1730 cm$^{-1}$. Note that the nominal PMMA peak intensity ratio of 2:1:2 is well reproduced in all spectra taken outside the antenna hotspots (red $\Delta T(\omega)$ curves).

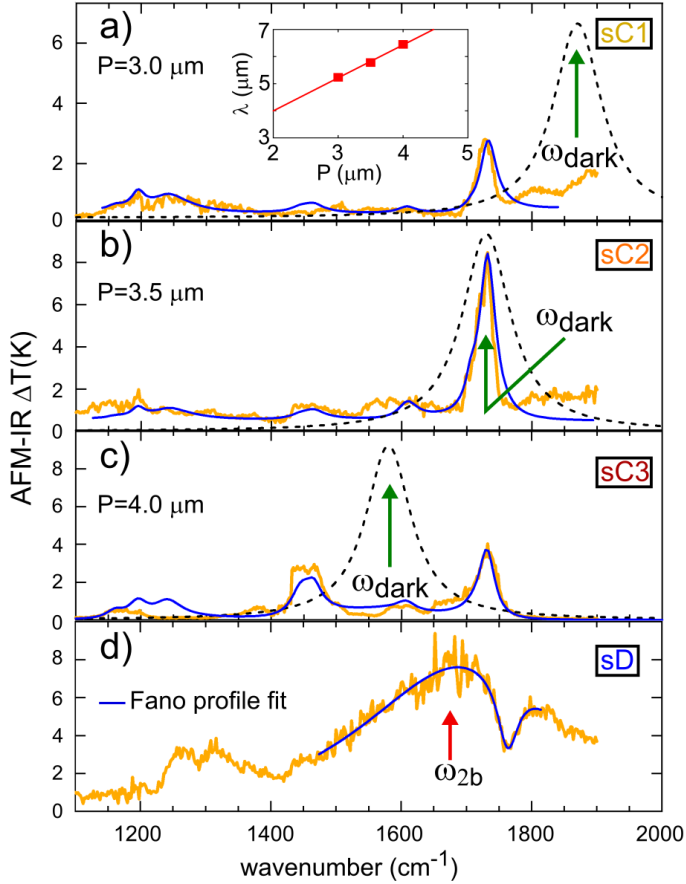

**Figure 4.** Surface plasmon-enhanced vibrational lineshapes as they appear in $\Delta T_{hs}(\omega)$ (orange curves). Blue curves represent the best-fit to the data of the respective model: a,b,c: thermoplasmonic SEIRA model for dark modes of Eq. 1; d) Fano model for the bright mode. The lorentzian lineshapes of the dark modes ($L_{dark}(\omega)$ in Eq. 1) are reported as black dashed curves with normalized intensities. Inset of panel a: linear dependence on $P$ of the dark mode resonant wavelength.

In Figures 4a-c, $\Delta T(\omega)$ for three samples with varying array period are reported. In samples sC1, sC2, sC3, the shorter $H = 2.2$ μm and the $\theta_{tilt} = 0°$ both contribute to shift the bright plasmonic mode at higher frequencies if compared to all other samples, out of the AFM-IR measurement range. The dark mode frequency, instead, is expected from simulations to be close to the PMMA vibration at 1730 cm$^{-1}$ and weakly tunable with $P$ (see Supporting Information). A strong enhancement of the PMMA vibration at 1730 cm$^{-1}$ is observed indeed in the $\Delta T(\omega)$ of Figures 4a-c. Ratios of 2:1:4, 2:1:14 and 2:6:8 for the relative PMMA vibrational peak intensities at 1250, 1450 and 1730 cm$^{-1}$ are observed in Figures 4a, 4b and 4c respectively, to be compared to the nominal ratio 2:1:2 of Figure 1g. SEIRA enhancements from two fold to seven-fold are then observed when the dark mode frequency resonates with the vibration of PMMA molecules in the antenna hotspots. In Figure 4d we show the different situation that arises in sample sD, where the second bright mode is centered at 1700 cm$^{-1}$ and partially overlaps with the strong vibration of PMMA at 1730 cm$^{-1}$. As it is well known [30-36], this occurrence leads to a characteristic Fano interference lineshape in the extinction spectra which, being a near-field interference phenomenon, has an absorptive counterpart that can be observed in $\Delta T(\omega)$ [47]. In other words, the dip in $\Delta T$ at 1730 cm$^{-1}$ is the result of destructive interference between the enhanced field of the plasmonic bright mode and the polarization field of the narrow PMMA vibration, which have a phase difference close to $\pi$ [38, 48]. As a result, the molecule absorption is not efficiently enhanced in the antenna hotspot, and the energy dissipation is smaller than in the absence of a molecular vibration: $\Delta T(\omega)$ in Figure 4a shows a peak at $\omega_{2b}$ = 1700 cm$^{-1}$, while it has a dip at 1755 cm$^{-1}$ and it is featureless at the bare PMMA vibration frequency at 1730 cm$^{-1}$. The bright mode lineshape is a quasi-continuum if compared to the narrow PMMA vibration, so the line profile of Figure 4d could be reproduced by a Fano model taken from Ref. [36] (blue curve). We recall that this destructive interference is the main reason why a true SEIRA phenomenon was seldom observed in mid-IR nanoantennas [30-36, 49], and the concepts of surface enhanced infrared scattering (SEIRS) and resonant SEIRA, have been developed instead [21].

At odds with bright modes producing destructive near-field interference when resonating with molecular vibrations (see Figure 4d), the dark mode produces a clear SEIRA effect where absorption of molecules in the hotspots is enhanced. Both these effects are here detected by AFM-IR working as a local absorption probe. The dark mode does not display a strong electrodynamic response to the incident wave, as seen from the almost complete absence of clear Joule heating peaks at $\omega_{\text{dark}}$ in $\Delta T(\omega)$ of Figures 4b and 4c. Joule heating is observed at the dark mode frequency in Figure 4a, although it gives smaller temperature increase if compared to the thermoplasmonic effect of SEIRA at 1730 cm$^{-1}$. Apparently, these broken-symmetry modes efficiently transfer the potential energy gained from the incident wave to the molecules located in the antenna hotspots, possibly due to a low fraction of free electron kinetic energy [50]. The dissipation of the potential energy of the dark mode absorbed by resonant molecule vibrations leads to the thermoplasmonic effect of SEIRA clearly observed in Figures 4b-d. For bright modes, where the distribution of potential and kinetic energy is more even, closely resembling a classical harmonic oscillator [51], destructive interference phenomena such as the one observed in Figure 4a prevent direct energy dissipation in the molecules (SEIRA), and the high fraction of kinetic energy favors Joule heating in the metal instead.

**Model**

In Figure 4a-c the blue curves superimposed to the orange curves represent a best fit to the data of a model described below for the spectral absorption lineshape $A(\omega)$ resulting from the e.m. resonance between the plasmonic dark mode and the mid-IR molecular vibrations. The model curve $A(\omega)$ can be fitted to the experimental $\Delta T(\omega)$ by using a frequency-independent multiplication factor accounting for thermal diffusion [24, 52]. Also included in the model is a constant background absorption $A_0$ related to laser heating of the cantilever, an effect specific to the AFM-IR technique. The total absorbance of e.m. energy from the incident wave, in the absence of Fano interference, is given by the sum of direct absorbance from all the molecules in the unit cell according to the Beer-Lambert's law (absorption coefficient $\alpha_{\text{mol}}$ times the polymer layer thickness $d_{\text{mol}}$, where the index "mol" runs over any type of molecule present in the sample) and of plasmon-enhanced absorbance in the antenna hotspot:

$$A(\omega) = A_0 + \sum_{\text{mol}} [d_{\text{mol}} \alpha_{\text{mol}}(\omega) + c_{\text{mol}} \alpha_{\text{mol}}(\omega) L_{\text{dark}}(\omega) r_{hs}] \qquad (1)$$

where $L_{\text{dark}}(\omega)$ is the intensity enhancement spectrum defined by the spectral response of the dark mode, $r_{hs}$ is a characteristic linear dimension of the hotspot, and $c_{\text{mol}}$ is the relative volume concentration of a given molecular species in the hotspot. In our experiment, the index "mol" runs only over the two polymers PMMA, AZ. The dark mode lineshapes $L_{\text{dark}}(\omega)$, assumed as Lorentizans (see Supporting Information) are plotted for different arrays in Figure 4b-d as dashed black curves. Importantly, the frequency selectivity of the thermoplasmonic effect of SEIRA implied by the model in Eq. (1), where Joule heating is neglected, corresponds to a *chemical* selectivity, through the substance specificity of mid-IR vibrational fingerprints. Comparing the $\Delta T$ spectrum in Figure 4a (conventional thermoplasmonic effect due to Joule heating by bright modes) with the spectra in Figure 4b-d (thermoplasmonic effect of SEIRA), it can be seen that the latter mechanism dominates at frequencies corresponding to narrow molecular vibrations resonating with dark modes, while the former mechanism dominates for bright modes.

**Evaluation of Temperature Increase and Gradient**

We now turn to the quantitative determination of the temperature increase in different locations of our samples for the different plasmonic modes. We focus on two types of locations: points of the sample surface just above the antenna hotspot (represented by orange pixels in the AFM-IR maps of Figures 1 and 4), reaching a high local temperature $T_{hs}(\omega)$ under laser illumination either at frequency $\omega_{2b}$ or at frequency $\omega_{\text{dark}}$, and points at the surface of the embedding layer at a distance $\sim P/2$ from the antenna hotspot (red pixels in Figures 1 and 4), reaching a global temperature $T_{\infty}(\omega)$ which may, or may not, be higher at the plasmonic mode frequencies than for any other frequency. The base environmental temperature in the absence of laser illumination is $T_{\text{env}} = 295$ K in all our experiments. In the resonantly-enhanced photothermal expansion version of AFM-IR used here [52], the equilibrium temperatures are reached within tens of nanoseconds from the onset of the laser pulse, which has a typical duration of 260 ns and a much longer repetition time of $\sim 6$ $\mu$s, so a steady-state equilibrium temperature is approximately reached during and after each laser pulse [25]. By following the procedure of Ref. [25], the value of the cantilever deflection signal induced by photoexpansion $V_{\text{PSPD}}$ can be used to calculate the temperature increases $\Delta T_{hs} = T_{hs} - T_{\text{env}}$ and $\Delta T_{\infty} = T_{\infty} - T_{\text{env}}$, which correspond to the values in the ordinate axes of $\Delta T(\omega)$ of the spectral plots in the Figures. The variation $\Delta \delta$ of the indentation depth $\delta$ during each laser pulse (photoexpansion length) is considered as the main mechanism of impulse transfer from the material to the cantilever through the probe tip. Considering the values of heat capacitance and thermal conductivity of polymers, a homogeneous temperature increase can be assumed for the layer contained in a cylinder extending between the antenna and the AFM tip apex, with height $h$ = 100 nm and circular base radius equal to the thermal diffusion length around 300 nm [23]. From the maximum experimental value of $V_{\text{PSPD}} \sim 1.0$ V, typically observed with laser power of 50 mW at either bright or dark mode frequencies, using the AFM-IR sensitivity value of ~20 mV/nm and the measured mechanical quality factor of our AFM-IR cantilevers of 113, we obtain $\Delta\delta \simeq 3{,}3 \cdot 10^{-11}$ m. Using a linear thermal expansion coefficient $\ell = 3 \cdot 10^{-5}$ K$^{-1}$ for PMMA we obtain a local $\Delta T_{\text{hs}} = \Delta\delta / (\ell h) = 11$ K for both bright and dark modes.

In order to compare our results with the predictions of existing thermoplasmonic models [53], at first-order approximation we can neglect the heat-sink effect of the substrate, which is very important for thin films and planar antennas [27], because the vertical antennas are almost fully embedded in relatively bad thermal conductors. We also assume instantaneous thermalization of each gold-coated antenna to the temperature reached in its hotspot. We can then adapt the formula from Ref. [53] to our experiment by inserting a spectral dependence of the absorbance $A(\omega)$ defined in Eq. (1):

$$\Delta T_{hs}(\omega) = \frac{I_0 A(\omega) d_{\text{ant}} r_{\text{hs}}}{4 R_{\text{eq}} \beta k_\infty} \qquad (4)$$

where $I_0$ is the peak laser power density normalized to 1.6 10$^8$ W/m$^2$ over the entire EC-QCL spectrum (see Methods), $A(\omega)$ is the absorbance spectrum of the given plasmonic mode, and $k_\infty$ is the thermal conductivity of the material that fills most of the space around the hotspot (here, the polymer AZ). The absorption cross-section of the dark mode is identified with the product $A(\omega) d_{\text{ant}} r_{\text{hs}}$, while $\beta$ and $R_{eq}$ are the thermal capacitance coefficient and equivalent radius introduced in Ref. [53] (in our case, $R_{eq}$ is 215 nm and $\beta = 1.2$ using the ellipsoid approximation for the vertical nanoantenna). The evaluation of Eq. (4) gives $\Delta T_{\text{hs}}(\omega_{\text{dark}}) \approx 10$ K, in agreement with the estimate of 11 K obtained from the AFM-IR signal calibration.

We now turn to the estimate of steady-state temperature gradients. The orange curves in Figure 3 represent the local temperature increase $\Delta T_{\text{hs}}(\omega)$. They show the apparent fact that bright and dark modes display comparable thermoplasmonic heat generation efficiency, although the heat generated by dark modes is mostly due to SEIRA and the heat generated by bright modes is entirely due to Joule effect. The red curves in Figure 3 can now be identified as $\Delta T_\infty(\omega)$ or the global temperature increase of the sample surface, given by the direct (Beer-Lambert) optical absorption of polymers, which can be rather strong in the mid-IR, plus an extra heat flux from the much hotter thermoplasmonic hotspots nearby. From the value of the red curves in Figure 3 at the strongest PMMA vibration frequency of 1730 cm$^{-1}$, we infer that the Beer-Lambert contribution to $\Delta T_\infty$ is $\leq 1$ K with focused mid-IR illumination of 50 mW and PMMA thickness of 100 nm. In order to estimate the contribution to $\Delta T_\infty$ due to the thermoplasmonic extra heat flux, we can use the small peak observed in the red curve of sample sA in Figure 3 $\Delta T_\infty(\omega)$ due to Joule heating at a bright mode frequency $\omega_{2b}$ far from any vibrational absorption: the peak intensity at $\Delta T_\infty(\omega_{2b})$ is an order of magnitude smaller than the peak intensity at $\Delta T_{\text{hs}}(\omega_{2b})$ (corresponding orange curve of sample sA in Figure 3). Since we have calculated $\Delta T_{\text{hs}}(\omega_{2b}) \approx 10$ K for 50 mW of laser power, we deduce that $\Delta T_\infty(\omega_{2b}) \approx 1$ K. Therefore $\Delta T_{\text{hs}} \approx 10$ K $\gg \Delta T_\infty \approx 1$ K and, considering a distance of 2 $\mu$m between the two locations, we conclude that there is a temperature gradient exceeding 5 K/$\mu$m at the plasmonic mode frequency of our structures at 50 mW of laser power, which could grow proportionally to 50 K/$\mu$m at the maximum available power of EC-QCLs around 500 mW.

We can compare the measured temperature gradients exceeding 5 K/$\mu$m with those predicted by the model of Ref. [53]. Therein, $\Delta T$ at a location distant $P/2$ from the thermoplasmonic heat source is obtained by substituting $P/2$ to $R_{\text{eq}}\beta$ in Eq. (4). Since $P/2$ is 1500 to 2000 nm in our structures, an order of magnitude larger than the product $R_{\text{eq}}\beta = 260$ nm. Therefore, a ratio around 10 is expected between $\Delta T_{\text{hs}}(\omega)$ and $\Delta T_\infty(\omega)$ for any $\omega$ corresponding to a plasmonic resonance, as found from the above analysis of the experiments: not only the absolute temperature increase, but also the gradient estimated from the AFM-IR spectra is in good agreement with thermoplasmonic models [53].

Given the obtained values of temperature increases and gradients and the maximum available power from commercial mid-IR EC-QCLs, we propose that vertical antenna arrays may be used as antenna-assisted thermoplasmonic surfaces for thermophoresis [18] or for laser-assisted catalysis of chemical reactions among molecular species immobilized on the surface of the antennas [16, 17]. The structures could also be exploited for global heating of macroscopic sample surfaces with narrowband laser light at the mid-IR vibrational resonance of given molecules. If compared to broadband mid-IR perfect absorbers, such as carbon black, nanoporous gold [36], or metasurfaces [15, 54, 55], the vertical antenna arrays presented in this work are substance-selective when operated at mid-IR wavelengths close to plasmonic dark modes, for which SEIRA is dominant over Joule heating. In other words, the thermoplasmonic effect of SEIRA operates at its maximum efficiency only if a target substance with a mid-IR fingerprint resonant with the dark mode is present in the antenna hotspots, otherwise the weak Joule heating effect of the dark mode alone, in the absence of SEIRA, is not efficient, effectively producing a kind of "mid-IR chemical trigger" of the thermoplasmonic temperature increase.

**Conclusions**

We have measured the thermoplasmonic response of vertical antenna arrays in the mid-infrared using a tip-enhanced photothermal expansion microscope to measure the local nanoscale temperature increase under illumination with a tunable-wavelength laser. The study of the temperature increase spectra measured with the scanning probe tip positioned above the antenna hotspot and away from the hotspots demonstrates the enhancement of the molecule vibration intensities when their mid-infrared fingerprints resonate with the plasmonic dark modes of the antenna arrays, here defined as modes with small scattering cross section. Due to the dominant non-radiative decay of molecule vibrations, their plasmon-enhanced intensity leads to surface-enhanced infrared absorption (SEIRA), which is directly measured as photoinduced temperature increase in our experiment. With 50 mW of focused mid-infrared laser illumination, local temperature increases of 10 K and strong temperature gradients above 5K/$\mu$m have been measured. The same enhancement of the molecule vibration intensity and absorption could not be obtained by illumination at the bright mode wavelengths, because of destructive Fano interference between the molecule polarization fields and the antenna near-fields. Local thermoplasmonic temperature increases up to 100 K should be possible by locating molecules with resonant mid-infrared fingerprints in the antenna hotspots and using focused beams from quantum cascade lasers at their typical full emission power of 500 mW.